\documentclass[useAMS,usenatbib]{mn2e}
\usepackage{graphicx}
\title{Kinetically Dominated FRII Radio Sources}
\author[Brian Punsly]{Brian Punsly \\
4014 Emerald Street No.116, Torrance CA, USA 90503 and \\
International Center for Relativistic Astrophysics,
I.C.R.A.,University of Rome La Sapienza, I-00185 Roma, Italy\\
E-mail: brian.m.punsly@L-3com.com or brian.punsly@gte.net}
\begin{document}
\maketitle \label{firstpage}
\begin{abstract}The existence of FR II objects that are kinetically dominated, the
jet kinetic luminosity, $Q$, is larger than the total thermal
luminosity (IR to X-ray) of the accretion flow, $L_{bol}$, is of
profound theoretical interest. Such objects are not expected in
most theoretical models of the central engine of radio loud AGN.
Thus, establishing such a class of objects is an important
diagnostic for filtering through the myriad of theoretical
possibilities. This paper attempts to establish a class of quasars
that have existed in a state of kinetic dominance, $R(t)\equiv
Q(t)/L_{bol}(t)>1$, at some epoch, $t$. It is argued that the 10
quasars in this article with a long term time average $Q(t)$,
$\overline{Q}$, that exceed $L_{Edd}$ are likely to have satisfied
the condition $R(t)>1$ either presently or in the past based on
the rarity of $L_{bol}>L_{Edd}$ quasars. Finally, the existence of
these sources is discussed in the context of the theory of the
central engine.
\end{abstract}
\section{Introduction}
The connection between accretion flow parameters and radio jet
power is mysterious. Most quasars are radio quiet, however
$\approx 10\%$ of optically selected quasars are radio loud and
more importantly about $2\%$ of quasars have powerful FRII radio
lobes \cite{dev06}. The lobe emission is the signature that the
time averaged jet kinetic luminosity, $\overline{Q}$, is enormous,
$\overline{Q}>10^{44} \mathrm{ergs/sec}$ \cite{pun01}. It is not
known how powerful the quasar jet can be relative to the thermal
luminosity from accretion, $L_{bol}$. Understanding the limits of
jet power can help reveal the physical nature of the quasar
central engine. The primary obstacle in this exercise is that in
order for $L_{bol}$ and $Q$ to be estimated contemporaneously
necessitates that $Q(t)$ be derived from parsec scale radio jet
observations and such efforts are plagued by poorly estimated
Doppler enhancement factors (raised to the fourth power) and the
results are often grossly inaccurate \cite{pun05,pun06}. For
example, the estimates of blazar jet power in \cite{cel97,wan04}
indicate that $R(t)\equiv Q(t)/L_{bol}(t)>1$ AGN are fairly
common, but the results are skewed by poorly constrained Doppler
factors \cite{tin05}. Alternatively, the time averaged jet power
$\overline{Q}$ can be estimated far more accurately from the
isotropic properties of the extended emission. Some
$\overline{R}\equiv \overline{Q}/L_{bol}>1$ sources were found in
\cite{pun06}. Unfortunately the $\overline{Q}$ estimate is not
contemporaneous with the $L_{bol}$ data, so one can not say if the
sources presently satisfy or ever satisfied $R(t)>1$. Grossly
inaccurate measurements of $Q$ are really of no value, so we must
concentrate on the $\overline{Q}$ estimates and combine this with
other information in order to establish a class of $R(t)>1$
sources. In section 3, a subsample of 10 FR II AGN with
$\overline{Q}_{Edd}\equiv \overline{Q}/L_{Edd}>1$, is argued to be
comprised of $R(t)>1$ sources.
\section{Estimation Techniques} In this section, we review the
standard estimation techniques used in the following analysis. The
more information that is known about the large scale radio
structure such as the radio spectral index across the lobe and
high resolution X-ray contours, the more sophisticated and
presumably more accurate an estimate that can be obtained for the
energy flux delivered to the lobes from the jet
\cite{pun05,bir04,pun01}. Unfortunately, such detailed information
does not exist for most radio sources and we need an expedience
that is helpful for studying large samples. Such a method that
allows one to convert 151 MHz flux densities, $F_{151}$ (measured
in Jy), into estimates of $\overline{Q}$ (measured in ergs/s), was
developed in \cite{wil99,blu00}, the result is captured by the
formula derived in \cite{pun05}:
\begin{eqnarray}
 && \overline{Q} \approx 1.1\times
10^{45}\left[X^{1+\alpha}Z^{2}F_{151}\right]^{\frac{6}{7}}\mathrm{ergs/sec}\;,\\
&& Z \equiv 3.31-(3.65)\times\nonumber \\
&&\left[X^{4}-0.203X^{3}+0.749X^{2}
+0.444X+0.205\right]^{-0.125}\;,
\end{eqnarray}
where $X\equiv 1+z$, $F_{151}$ is the total optically thin flux
density from the lobes (i.e., \textbf{contributions from Doppler
boosted jets or radio cores are removed}). In this paper we adopt
the following cosmological parameters: $H_{0}$=70 km/s/Mpc,
$\Omega_{\Lambda}=0.7$ and $\Omega_{m}=0.3$.  We define the radio
spectral index, $\alpha$, as $F_{\nu}\propto\nu^{-\alpha}$. The
formula is most accurate for large classical double radio sources,
thus we do not consider sources with a linear size of less than 20
kpc. Alternatively, one can also use the independently derived
isotropic estimator from \citet{pun05}
\begin{eqnarray}
&&\overline{Q}\approx
5.7\times10^{44}(1+z)^{1+\alpha}Z^{2}F_{151}\,\mathrm{ergs/sec}\nonumber\\
&& \quad\alpha\approx 1\;.
\end{eqnarray}
\par Even though one can use the
jet emission from the parsec scale radio core to estimate, $Q$
more contemporaneously with the accretion flow emission as in
\cite{cel97}, such estimates are prone to be very inaccurate. One
is observing a very small amount of "waste energy" such as X-ray
or optical emission as the powerful radio jet propagates away from
the source. One must then try to figure out the fraction of $Q$
that is converted to waste energy in this region. Typically, the
X-ray, radio and optical regions are observed with different
spatial resolution, so it is unclear if one is detecting the same
physical region on parsec scales as one synthesizes the broad band
data. In cases in which there is sufficient broad band flux to
make an estimate, one is plagued with the further ambiguity of
determining the Doppler factor of the relativistic jet. This is a
critical obstacle because the luminosity from an unresolved region
scales with the Doppler factor to the fourth power \cite{lin85}.
More shortcomings of this method are discussed in \cite{tin05} and
a published example in which the $Q$ is apparently miscalculated
using radio core properties by three orders of magnitude is
discussed explicitly. Thus, the most accurate estimates of $Q$
should use an isotropic estimator such as the radio lobe flux in
(1), even though it is just a time average.
 \par The total
bolometric luminosity of the accretion flow, $L_{bol}$, is the
thermal emission from the accretion flow (IR to X-ray), including
any radiation in broad emission lines. To estimate, $L_{bol}$, we
use the composite SED in table 2 of \cite{pun06}, which see for
details. This spectrum, in combination with
  the broad emission lines, represents the ``typical'' radiative
signature of a strong accretion flow onto a black hole. This
signature is empirical and it is
  independent of all theoretical models of the accretion flow. If $L(\nu)_{\mathrm{obs}}$
  is the observed spectral luminosity at the quasar rest frame frequency,
  $\nu$, then $L_{bol}$ is estimated as
\begin{eqnarray}
&& L_{bol}=1.35\frac{\nu L(\nu)_{\mathrm{obs}}}{\nu
L(\nu)_{\mathrm{com}}}\times 10^{46}\mathrm{ergs/sec}\;,
\end{eqnarray}
 where $L(\nu)_{\mathrm{com}}$ is the spectral luminosity from the composite SED.
\par Finally, we
list the equations used to estimate the central black hole mass,
$M_{bh}$. The reader should consult the references for details. We
estimate $M_{bh}$ from the line widths, F(H $\beta$) or F(C IV),
\cite{ves06} or F(Mg II), \cite{kon06}, where F() means "FWHM of"
\begin{eqnarray}
 && M_{bh}(H \beta)= \nonumber \\
 && 10^{6.91 \pm
0.02} \left[\left(\frac{F(H\beta)}{1000
\mathrm{km/s}}\right)^{2}\left(\frac{\lambda L_{\lambda}(5100
\AA)}{10^{44}\mathrm{ergs/s}}\right)^{0.50} \right]\;,\\
&& \nonumber \\
&& M_{bh}(\mathrm{CIV})= \nonumber \\
&& 10^{6.66 \pm 0.01}\left[\left(\frac{F(\mathrm{CIV})}{1000
\mathrm{km/s}}\right)^{2}\left(\frac{\lambda L_{\lambda}(1350
\AA)}{10^{44}\mathrm{ergs/s}}\right)^{0.53} \right]\;,\\
&& \nonumber \\
&& M_{bh}(\mathrm{MgII})= \nonumber \\
&& 10^{6.53} \left[\left(\frac{F(\mathrm{MgII})}{1000
\mathrm{km/s}}\right)^{2}\left(\frac{\lambda L_{\lambda}(3000
\AA)}{10^{44}\mathrm{ergs/s}}\right)^{0.58 \pm 0.10} \right]\;.
\end{eqnarray}

\begin{table*}
 \centering
\caption{FR II Quasars with Super Eddington Jets} {\footnotesize
\begin{tabular}{ccccccccc}
 \hline

Source &   $z$ &  $\overline{Q}$& $L_{bol}$ &  $\overline{R}$  &  freq & $L_{bol}/L_{Edd}$ & $\overline{Q}_{Edd}$ &  ref\\
       &    &  $10^{45}\mathrm{ergs/s}$ & $10^{45}\mathrm{ergs/s}$  &  & ($10^{15}$ Hz) &  & &\\
\hline 3C 216   & 0.670 &  15.1/14.1  & $\approx 0.12$ & $\approx
120$ &
0.71/1.16  & $0.05 - 0.1$ & $3.3-10$ & 1 \\
3C 455 & 0.543 &  7.13/5.04 & $0.38$ &  $ 18.7/13.3 $   & 0.94 &
$1.42$ & $26.7/18.9$& 2 \\
   &  &  7.13/5.04 & $0.38$ &  $ 18.7/13.3 $   & 0.94 &
$0.07$ & $1.33/0.94$& 3 \\
 3C 82      &  2.878 &  155.4/183.8  & 14.5 &  10.7/12.7 & 0.014 & 0.106 & 1.14/1.35 & 4 \\
          &   &  155.4/183.8  & 25.0 &  6.22/7.35 & 1.67  & 0.245 & 1.52/1.80 &  4 \\
 3C 9      &  2.009 &  148.3/174  & 25.0 &  5.93/6.96    & 1.67  & 0.264 &1.57/1.85 & 5 \\
          &  &  148.3/174  &  38.8 & 3.82/4.49    & 0.0078  & 0.324 & 1.24/1.46& 6 \\
 4C 25.21  & 2.686 &    59.3/59.7  & 11.6 &  5.11/5.15    & 1.14  & 0.198 & 1.02/1.02 &  5 \\
 PKS 1018-42  & 1.28 &  63.9/65.2  & 19.3 &  3.31/3.38    & 1.37 & 0.428 & 1.42/1.45 & 7 \\
             &  &  63.9/65.2  & 14.7 &  4.35/4.45    & 1.37  & 0.326 & 1.42/1.45 &  7 \\
 4C 04.81  & 2.594 &  103.8/148  & 35.8 &  2.90/4.13    & 2.30  & 0.459 & 1.33/1.90 &  5 \\
 3C 196   & 0.871 &  73.5/87.0  & 31.6 &  2.33/2.76   & 1.53  & 3.04 & 7.10/8.41 & 8 \\
         &  &  73.5/87.0  & 31.6 &  2.33/2.76   & 1.53 & 0.238 & 0.66/0.56 & 9 \\
 3C 14   & 1.469 &  52.38/51.68  & 32.6 &  1.61/1.59   & 1.00  & 0.604 & 1.05/1.03 & 10 \\
 3C 270.1   & 1.519 &  65.1/66.6  & 48.2 &  1.35/1.38   & 2.07  & 0.844 & 1.14/1.17 & 5 \\

\hline
\end{tabular}}
\par
\footnotesize{1. see \cite{pun07}, 2. continuum and FWHM from
\cite{gel94}, $M_{bh}$ from eqn (5), 3. $M_{bh}$ from bulge
luminosity estimate in eqn (8), 4. $L_{bol}$ and FWHM raw data
from \cite{sem04}, $M_{bh}$ from eqn (6), 5. $L_{bol}$ and FWHM
from \cite{bar90}, $M_{bh}$ from eqn (6), 6. continuum from
\cite{mei01}, FWHM from \cite{bar90},$M_{bh}$ from eqn(6), 7.
\cite{pun06}, $M_{bh}$ from eqn(7), 8. continuum and FWHM from
\citet{law96}, $M_{bh}$ from eqn (5), 9. continuum and FWHM from
\cite{law96}, $M_{bh}$ from eqn (7), 10. continuum and FWHM from
\cite{aar05}, $M_{bh}$ from eqn (7)}
\end{table*}

\section{Kinetically Dominated Sources} Table 1 is comprised of $\overline{Q}_{Edd}\equiv
\overline{Q}/L_{Edd}>1$ QSOs that were found by cross-correlating
all existing VLA, MERLIN and ATCA radio maps with the absolute
visual magnitudes and broad line widths of the QSOs in
\cite{ver91}. Sources that were discovered by this method with
$\overline{R}>4$ were previously reported in \cite{pun06}, which
see for more details on the sample selection. A more in depth
analysis showed that 3C 216 and 3C 455 also belong in the table.
However, as mentioned previously, due to the lack of simultaneity
between jet ejection and UV emission, the \cite{pun06} sample does
not establish the $R(t)>1$ condition. In subsection (3.1) it is
argued that the $\overline{Q}/L_{Edd}>1$ jets in table 1 satisfy
$R(t)>1$, for some $t$. The first column is the source name
followed by the redshift. Column 3 has two estimates of
$\overline{Q}$ from (1) on the left and (3) on the right,
separated by a slash. Column 4 is $L_{bol}$ estimated from (4) at
the rest frame frequency in column 6. Column 5 is $\overline{R}$.
The last three columns are $L_{bol}/L_{Edd}$, $\overline{Q}_{Edd}$
and the references. If more than one observed frequency was
available, a second estimate for $L_{bol}$ was provided for that
source so that one can assess the error in using a composite SED
to approximate $L_{bol}$.
\par The QSO, 3C 455, requires some elaboration. This object has a narrow
H$\beta$ (FWHM = 620 km/s), yet the nuclear luminosity is far
stronger than a narrow line radio galaxy and its magnitude is more
typical of a strong Seyfert 1 galaxy, hence its historical
classification as a quasar \cite{gel94}. Implicit in the first
estimate in table 1 is that the object is a narrow line Seyfert 1,
but H$\beta$ appears weak compared to [OIII]$\lambda 5007$
(H$\beta$/[OIII]=0.2 and H$\beta$/[OIII]$>0.33$ in narrow line
Seyfert1's) because the narrow line region is greatly enhanced by
emission line gas that is excited by the jet and is aligned with
the radio structure \cite{dev99}. In support of this
interpretation, there is an elongated patch of diffuse emission
enveloping the kpc radio jet axis between the two lowest contours
of the HST image in \cite{leh99} with a flux approximately equal
to the narrow line flux in the F702W filter \cite{gel94,dev99}.
Alternatively, one can interpret the object as a narrow line radio
galaxy and $M_{bh}$ can be estimated from the galactic host bulge
luminosity. There are a variety of fits to the $M_{bh}$ - host
bulge luminosity relation, but the fits of \cite{dun02} offer the
smallest scatter since they are based on large samples. The
relation that is most relevant is the best fit estimator derived
from 72 AGN in table 3 of \cite{dun02},
\begin{eqnarray}
&& \log(M_{bh}/M_{\odot})= (-0.46\pm 0.03)M_{R} -(2.55\pm 0.72)\,.
\end{eqnarray} Thus, we need to find $M_{R}$ for the galactic bulge in
order to utilize (8). The HST image is taken with the F702W filter
and the apparent magnitude of the galactic bulge (after
subtraction of the nuclear core and extended narrow line flux) is
$m_{702W}=21.53$ \cite{dev99}. The $m_{702W}$ magnitudes are
approximately the Cousins R magnitudes. The transformation to the
Cousins R magnitude along with the k-correction that is given in
\cite{fuk95} yields $M_{R}=-21.5$ and from (8), $M_{bh}=4.2 \times
10^{7} M_{\odot}$. This is the basis for the second estimate in
table 1.
\subsection{Super Eddington Jets}
A value of $\overline{Q}_{Edd}>1$ suggests that even if the jet
central engine is currently in a low state then at some epoch
during the lifetime of the source it must have had $R(t)>1$. For
example, if PKS 1018-42 is not now nor has ever been kinetically
dominated then the on average $\overline{L}_{bol}>1.45 L_{Edd}$
for the lifetime of the source. However, since
$\overline{Q}=(1/T)\int_{0}^{T}Q(t) dt$, the peak values of the
instantaneous $Q(t)$, $\mathrm{max}_{t}\left[Q(t)\right]>1.45
L_{Edd}$. Thus, one expects that $\mathrm{max}_{t}\left[
L_{bol}(t)\right]$ would have to exceed $1.45 L_{Edd}$ by a
significant amount at certain epochs in order for the PKS 1018-42
to have always been in an $R(t)<1$ state. This is inconsistent
with the magnitudes of the peaks in the duty cycle of quasar
$L_{bol}(t)/L_{Edd}$ based on our current knowledge, see
\cite{bor02,lho05}. More importantly, the estimates of 268 AGN in
\cite{pun07} utilized the same estimators, eqns. (5)-(7), used
here and $L_{bol}/L_{Edd}<1$ for all the broad line AGN.
Furthermore, note that most of the QSOs in table 1 have prodigious
accretion rates and $\overline{R}>1$ is not a consequence of the
quasars being in a state of suppressed accretion. The most likely
explanation of the data in table 1 is that these sources
experienced epochs in which $R(t)\sim 1-10$ as opposed to the
alternative explanation that the broad line sources had protracted
phases in which $L_{bol}(t)/L_{Edd}\sim 1-10$.

\subsection{The Validity of the Estimates}Since the data set is small, the analysis
of the data is particularly sensitive to the integrity of the
estimates in table 1. The main drawback to the argument above is
that the estimates of $M_{bh}$ could be off by a large amount in
an individual object, \cite{ves06}, and these sources appear to
have $\overline{Q}_{Edd}>1$ merely as a result of these errors. If
this were the fundamental explanation of the
$\overline{Q}_{Edd}>1$ estimates then a statistical anomaly must
also  present, the $0.5\leq \overline{Q}_{Edd}\leq 1.0$ QSOs just
below the threshold of table 1 never have $M_{bh}$ over estimated
(i.e. there are no false negatives for the criteria
$\overline{Q}_{Edd}>1$). In subsection 3.2.3, it is shown that
this anomalous requirement conflicts with the data used to create
the virial mass estimates. Furthermore, there could also be errors
of smaller magnitude associated with the estimates of
$\overline{Q}$ \cite{pun05}.
\subsubsection{Suppressing Error Propagation} The possibility of these types of errors in table 1
is dealt with by introducing as many different sets of
observational data as possible in order to obtain independent
estimates of the same quantities. Thus, any results that are way
"out of family" like $\overline{Q}_{Edd}=26.7$ for 3C 455 can be
flagged as unreliable. In order to expose the largest potential
source of error, $M_{bh}$, is computed from as many different
spectral bands and emission lines as possible (generally two
entries in table 1). Similarly, we compute $\overline{Q}$ from two
different estimators. Furthermore, we estimate all potentially
reddened QSOs by finding an IR flux estimate to reveal hidden AGN
UV luminosity (this will affect (5)-(7)). For example, 3C 190
fails to meet the $\overline{Q}_{Edd}>1$ criteria because it has
considerable hidden AGN UV emission, see table 3 of \cite{pun06}.
The errors might still persist and the potential effects are
discussed in the following two subsections.
\begin{figure}
\includegraphics[width=85 mm]{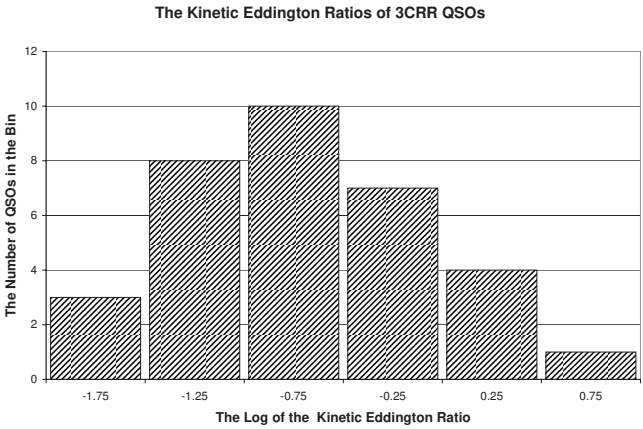}
 \caption{A histogram of the distribution of $\log
(\overline{Q}_{Edd})$ for the 3CRR QSOs $\geq 20$ kpc in linear
extent. The center of each bin is indicated on the horizontal
axis. All of the sources in the two bin on the far right satisfy
$\overline{Q}_{Edd}>1$. The mean value of
$\overline{Q}_{Edd}=0.52\pm 0.9$}
    \end{figure}
\subsubsection{3CRR Quasars} One way to assess the role of the statistical
 scatter in the estimators is to study the
complete 3CRR sample. We estimated $\overline{Q}_{Edd}$ for every
QSO $>20$ kpc (remember, (1) and (3) are not accurate for linear
sizes less than 20 kpc). Figure 1 is a histogram of the
distribution of $\log (\overline{Q}_{Edd})$. Every effort was made
to average away errant estimates. Every available broad line FWHM,
\cite{law96,bro94,wil95,aar05,guu01,mar03,bar90,kur04},  and
continuum flux density,
\cite{law96,wil95,aar05,bar90,kur04,mei01,sim00}, was gathered for
each source from the literature and every combination of these was
used to estimate numerous values of $M_{bh}$ from (5)-(7). These
values were then averaged for each source. This average $M_{bh}$
and the average $\overline{Q}$ from (1) and (3) were used to
compute the "best estimate" of $\overline{Q}_{Edd}$ for each
source. These estimates were then assembled in the histogram of
figure 1. This histogram shows that the $\overline{Q}_{Edd}>1$
QSOs are just the high end of a smooth distribution of
$\overline{Q}_{Edd}$ for the complete sample and are not outliers.
Over 30\% of the QSOs have $\overline{Q}_{Edd}>0.5$. This method
of assigning $\overline{Q}_{Edd}>1$ to a QSO is very conservative.
A single large estimate for $M_{bh}$ will swamp the other smaller
$M_{bh}$ estimates and suppress the large $\overline{Q}_{Edd}$
values (such as in 3C 196, 3C 190 and 3C 455). If one were to
average the $\overline{Q}_{Edd}$ values instead then large
$M_{bh}$ would be suppressed and there would be four 3CRR QSO with
$\overline{Q}_{Edd}>4$.

\subsubsection{An Analysis of the Distribution of Errors} There are two major
facts that allow a clear interpretation of the effect of the
errors in the estimation techniques on the $\overline{Q}_{Edd}\sim
1$ QSOs in table 1 and figure 1. Firstly, the detailed
investigation of 3C 216 in \cite{pun07} illustrates the existence
of $\overline{Q}_{Edd}>1$ QSOs very convincingly. The
$\overline{Q}_{Edd}>3.3$ condition in 3C 216 is verified by three
independent estimators of $M_{bh}$, F(H$\beta$), F(Mg II), and the
host bulge luminosity. Secondly, there is no evidence that the
scatter in the best fit estimators is skewed asymmetrically to low
values of $M_{bh}$ \cite{ves06,kon06}. Comparisons to
reverberation mapping based estimates indicate that the errors are
symmetrically distributed about the best fit estimators, eg.
figure 9 of \cite{ves06}. If not for these facts, one could take a
pessimistic view and plausibly argue that the
$\overline{Q}_{Edd}>1$ sources in table 1 and figure 1 are an
artifact of selecting those sources in which the errors associated
with the line width estimation techniques have underestimated
$M_{bh}$. This argument implicitly assumes that the high end of
the $\log(\overline{Q}_{Edd})$ distribution represents the
variance in the line width estimators, not the actual spread in
$M_{bh}$. However, the example of 3C 216 demonstrates that
$\overline{Q}_{Edd}>1$ sources exist, indicating that the more
reasonable explanation of figure 1 is that the errors in the
estimates are randomly distributed and they are not skewed
preferentially to produce low $M_{bh}$. In particular, some of the
sources with $\overline{Q}_{Edd}>1$ are probably actually,
$\overline{Q}_{Edd}<1$ sources, but a similar number of the
$0.5<\overline{Q}_{Edd}<1$ sources are actually
$\overline{Q}_{Edd}>1$ sources.
\par It is worth expanding the
discussion of the $\overline{Q}_{Edd}>1$ sources beyond the most
conservative of claims made above. This condition is far stronger
than $Q(t)_{\mathrm{max}}>Q_{Edd}$ and in general indicates
episodic values of $Q(t)$ considerably in excess of $Q_{Edd}$.
More realistically based on the discussion of section 3.1, all the
sources that have $\overline{Q}_{Edd}>0.5$ almost certainly had
episodes in which $Q_{Edd}>1$ and the $R(t)>1$ condition was
satisfied. Considering that the median value in figure 1 is
$\overline{Q}_{Edd}=0.26$, $\sim 0.3 - 0.5$ of the the 3C sources
were likely to have been kinetically dominated at some time in
their past. This claim can not be extended to the FR II population
as a whole, but is a consequence of the fact that the 3C QSOs
typically reside at the high end of the steep extended luminosity
distribution for radio loud quasars.

\section{Discussion} The article demonstrates that many FR II
sources are likely to exist in a state of $R(t)>1$. The
$\overline{Q}_{Edd}>1$ jets must satisfy
$\mathrm{max}_{t}\left[R(t)\right]>1$, or else the accretion disks
of the broad line AGN would episodically be in an un-physical
state of $L_{bol}(t)/L_{Edd}\gg 1$. The Mid-IR 3C sample of
\cite{ogl06} indicated that $\sim 1/2$ of the FR II narrow line
radio galaxies had no hidden quasar. Based on the discussion
above, these are likely candidates to be in a state of $R(t)>1$.
\par As a consequence of the near independence of the UV spectrum on the radio state,
\cite{dev06,cor94}, it has been argued that the black hole and not
the accretion disk is the power source for FR II jets
\cite{sem04}. A large scale magnetic flux trapped within the
central vortex of an accretion disk can produce relativistic jets,
as seen in the MHD numerical simulations of
\cite{haw06,mck05,gam04}. Yet, as discussed in \cite{pun07} these
simulations are restricted to a maximum $R$ value, $R_{max}<0.5$.
The magnetic field line configuration leading to a relativistic
jet is created and maintained by "pinning" the poloidal flux to
the event horizon by a strong accretion flow in conjunction with
an incredibly massive and unobserved enveloping wind that
transports $>1 M_{\odot}/\mathrm{yr}$ of relativistically hot
protonic gas at 0.3c-0.4c for an FR II quasar \cite{pun07}. The
black hole actually gains energy over time due to the intense
accretion flow (and associated energy influx) required to compress
a large magnetic flux onto the small surface area of the event
horizon. Thus, a jet with large Q requires an intense accretion
flow that will radiate profusely for the radiative efficiency
predicted in these simulations, \cite{bek06}, and
$R_{\mathrm{max}} \sim 0.1$

\par Alternatively, the black hole energy extraction model of
\cite{pun01} and references therein that was numerically realized in
\cite{sem04} is based on large scale magnetic flux that threads the
equatorial plane of the ergosphere (the active region) of a black
hole \cite{pen69}. The theoretical $Q$ values are 2 orders of
magnitude larger than for flux pinned on the event horizon because
the surface area of the equatorial plane in the ergosphere is $\sim
10$ times larger than the surface area of the horizon that is
threaded by magnetic flux in simulations of \cite{haw06,mck05,gam04}
for rapidly spinning black holes (parameterized by $a/M\approx 1$,
where "a" is the angular momentum per unit mass of the black hole in
geometrized units), and the jet power scales like the surface area
squared \cite{sem04}. These solutions attain a value,
$R_{\mathrm{max}}$, that was calculated in \cite{pun07,pun01},
\begin{eqnarray}
R_{\mathrm{max}}\approx 11\frac{0.3}{\epsilon}(a/M)^{10.5}\; ,\,
0.90 <a/M<0.99\;,
\end{eqnarray}
where $\epsilon$ is the radiative efficiency of the accretion
flow, $L_{bol}\equiv \epsilon (dM/dt)$ and $(dM/dt)$ is the
accretion rate from the disk to the black hole. The maximum
efficiency from a thin disk is $\epsilon\approx 0.3$ \cite{nov73}.
For high $L_{bol}$ objects, luminous quasars, one expects
$\epsilon$ to be near maximal and $a/M\approx 1$ (see
\cite{bar70,elv02}), thus $R_{\mathrm{max}}\sim 10$ which would
explain the high Q episodes required to maintain a long term time
average of $\overline{Q}_{Edd}>1$.
\par These arguments seem to indicate that the essential element that
separates a radio quiet QSO from one that launches a strong FR II
jet with $R\sim 1$ is the formation of strong vertical magnetic flux
in the equatorial plane of the ergosphere. This is prevented in the
simulations of \cite{haw06,mck05,gam04} as a very weak seed magnetic
field is swept up in a single MHD fluid and compressed onto the
black hole. Alternatively, \cite{spr05} have investigated low
angular momentum flux tubes that evolve separately from the bulk of
the MHD fluid. In the two-fluid dynamics, the strong flux tubes
"swim" relative to the bulk MHD fluid and slowly move into the
ergosphere. This could be an essential piece of missing dynamics in
the present codes.

\end{document}